\documentclass[journal]{IEEEtran}
\ifCLASSINFOpdf
\else
\fi
\usepackage[dvips]{graphicx}
\begin{document}
\title{Comparison of Loss ratios of different scheduling algorithms}

\author{Sudipta~Das,~
        Lawrence~Jenkins~
        and~Debasis~Sengupta.
\thanks{S. Das and L. Jenkins are with the Department
of Electrical Engineering, Indian Institute of Science, Bengaluru, India.}
\thanks{D. Sengupta is with the Indian Statistical Institute, Kolkata.}
}


\maketitle

\begin{abstract}
It is well known that in a firm real time system with a renewal arrival process, exponential service times and independent and identically distributed deadlines till the end of service of a job, the earliest deadline first (EDF) scheduling policy has smaller loss ratio (expected fraction of jobs, not completed) than any other service time independent scheduling policy, including the first come first served (FCFS). Various modifications to the EDF and FCFS policies have been proposed in the literature, with a view to improving performance. In this article, we compare the loss ratios of these two policies along with some of the said modifications, as well as their counterparts with deterministic deadlines. The results include some formal inequalities and some counter-examples to establish non-existence of an order. A few relations involving loss ratios are posed as conjectures, and simulation results in support of these are reported. These results lead to a complete picture of dominance and non-dominance relations between pairs of scheduling policies, in terms of loss ratios.
\end{abstract}


%
\IEEEpeerreviewmaketitle

\section{Introduction}
\IEEEPARstart{I}{n} real time systems consisting of aperiodic jobs, such as web server, network router or real time database; it is typically not known when the job will arrive or what will its service time and its deadline be. If too many jobs arrive simultaneously, the system becomes overloaded and the jobs begin to miss their deadlines. The service requirements for the jobs are often not known beforehand, and hence are specified in probabilistic terms. So a fundamental problem in such systems is to schedule a set of jobs such as to allow the maximum possible number of jobs to meet their respective deadlines.

In this article, we consider various scheduling algorithms for firm real time systems (i.e., systems where a job must leave the queue after its deadline \cite{firm}) with a single processor and an aperiodic workload, under their commonly used model as a single server queue with an infinite buffer \cite{mom}. The infinite buffer ensures that there is no upper limit on the maximum number of jobs that can remain in the system. We adopt the usual assumption that every job is ready as soon as it is released, it can be preempted at any time and it never suspends itself. Moreover, we assume that the deadline of a job is till the end of its service, that the context switch overhead is negligibly small when compared with the service times of the tasks, and that the number of priority levels is unlimited. We also assume that job arrivals follow a renewal process with rate $\lambda$ (i.e., the mean inter-renewal time is $1/\lambda$), the service times are independent and identically distributed random variables with mean $1/\mu$, and the relative deadlines are also independent and identically distributed random variables with mean $1/\delta$. The service time and absolute deadline of a job are assumed to be known at its arrival epoch. For example, in a Web server, the name of the requested URL can be used to look up the length of the requested page and, hence, estimate the service time of the request.

In the above set up, the simplest scheduling policy is the First Come First Served ({\it FCFS}) policy, which stipulates that jobs be serviced in the order of their arrival. A more complex scheduling policy that has some attractive optimality properties is the Earliest Deadline First ({\it EDF}) policy \cite{rtsmom}. According to this policy, jobs that have arrived and await execution are kept in a ready queue, sorted in ascending order by their absolute deadlines. When the processor finishes a job, the first job in the queue is selected for execution. When a job arrives, it is inserted in the proper position of the queue (breaking ties arbitrarily). A variant of the \textit{EDF} policy provides for preemption of the currently running job by a newly arrived job, if the absolute deadline of this job is earlier than that of the currently running job. If it is assumed that a job can always be preempted, and that there is no cost of preemption, then it can be  shown that preemptive {\it EDF} is the optimal policy within the class of non-idling service-time independent preemptive policies \cite{opti_preemptive}, i.e., {\it EDF} can produce a feasible schedule of a set of jobs J with arbitrary release times and deadlines on a processor iff J has feasible schedule. Also, it has been shown that {\it EDF} stochastically minimizes the loss ratio (i.e., the fraction of jobs not completed) in both preemptive and non-preemptive models \cite{opti_des,opti_des_c}.

There have been attempts to reduce the loss ratio by controlling admission of newly arriving jobs in the queue, through a scheduling test. Prominent examples of this innovation are utilization based admission controller \cite{utb1} and the exact admission controller \cite{rtsmom}. The Utilization based admission-controller for aperiodic jobs is pessimistic in the sense that it sometimes denies admission to a job even if that job can be scheduled at that instant. It can be shown that a utilization based admission-controller also passes some jobs that would not be completed before their respective deadlines. The exact admission controller ({\it EAC}) seeks to remove these shortcomings at the cost of increased computational complexity  (\textit{O}(log$n$) for EAC as opposed to \textit{O}(1) for the utilization based admission controller) \cite{eac}.

While an admission controller takes into account the history of jobs already in the queue, a particular decision regarding admission may appear to be unduly conservative in the light of events that follow that decision. If the decision to serve a job is deferred till the epoch of it being served, then that decision can be made on the basis of additional information. Here, we consider a simple modification to scheduling protocols, called the early job discarding ({\it EDT}) technique. The  \textit{EDT} does not check the scheduling feasibility of a job on its arrival, but rather admits each incoming job into the system, inserts the job in an appropriate place of the queue according to the protocol being used and lets the system evolve. It discards a job at the epoch of its getting the server from the head of the queue, irrespective of it being a fresh job or a previously preempted job requesting the server again, if it is clear at that moment that the job cannot be completed before the deadline. The name early job discarding technique reflects the fact that it discards a job \textit{before} its deadline epoch. It should be noted that \textit{EDT} may not be feasible in applications that demand guaranteed completion of jobs once they are admitted to the queue. On the other hand, even where it is feasible, the value of this common sense belt-tightening step in improving the performance of a scheduling policy has never been formally studied.\footnote{The only relevant work that we could access in this connection is a simulation study in \cite{utb1}, where it was found that EDT works marginally better than the utilization based admission controller in a particular situation.}  We show that this step can be more effective than admission controllers in cutting down the loss ratio. 

In this article, we compare the performances of different scheduling strategies in terms of the job loss ratio. We show that, under a purely random environment, the inclusion of EDT or EAC in the FCFS and EDF scheduling policy reduces the loss ratio. We also prove that {\it EDF} along with {\it EDT} has smaller loss ratio than all other scheduling algorithms considered here.

This article is organized as follows. In Section \ref{dom_nondom}, possible dominance relations of the scheduling strategies in terms of loss ratio are discussed. Special attention to systems with deterministic job deadlines is given in Section \ref{degenerate}. Some concluding remarks are provided in Section \ref{conclusion}, while proofs of all the result are presented in the appendix.

\section{Some dominances and non-dominances} \label{dom_nondom}
Let $\alpha_{sp }^{G}$	denote the loss ratio of the system under scheduling policy \textit{sp} and relative deadline distribution \textit{G}. The following proposition follows from Theorem~1 of Towsley and Panwar \cite{opti_des}.

\textit{Proposition} 1. 
In an $G/M/1/G$ queue, the loss ratio under the \textit{EDF} scheduling policy is smaller than that under any other service time independent scheduling policy with deadline till the end of service.

In particular, \textit{EDF} produces smaller loss ratio than \textit{FCFS}, i.e., $\alpha_{EDF }^{G}\le\alpha_{FCFS }^{G}$.

In this section, we look for similar dominance relations between pairs of the scheduling policies \textit{FCFS}, \textit{EDF}, \textit{FCFS-EDT} (FCFS along with EDT), \textit{EDF-EDT} (EDF along with EDT), \textit{FCFS-EAC} (FCFS along with EAC) and \textit{EDF-EAC} (EDF along with EAC). 

\bigskip
\textit{Proposition} 2. 
	 In a $G/G/1/G$ queue, the loss ratio under the \textit{EDF} scheduling policy can only reduce when Early Discarding Technique is used, i.e., $\alpha_{EDF\mbox{-}EDT }^{G} \leq \alpha_{EDF }^{G}$.
	
\bigskip
\textit{Proposition} 3. 
	 In a $G/G/1/G$ queue, the loss ratio under the \textit{FCFS} scheduling policy can only reduce when Early Discarding Technique is used, i.e., $\alpha_{FCFS\mbox{-}EDT }^{G} \leq \alpha_{FCFS }^{G}$.
	
\bigskip
\textit{Proposition} 4. 
	 In a $G/G/1/G$ queue, the loss ratio under the \textit{EDF} scheduling policy can only reduce when Exact Admission Control is used, i.e., $\alpha_{EDF\mbox{-}EAC }^{G} \leq \alpha_{EDF }^{G}$.
	 
\bigskip
\textit{Proposition} 5. 
	 In a $G/G/1/G$ queue, the loss ratio under the \textit{FCFS} scheduling policy can only reduce when Exact Admission Control is used, i.e., $\alpha_{FCFS\mbox{-}EAC }^{G} \leq \alpha_{FCFS }^{G}$.

\bigskip
\textit{Proposition} 6. 
	 In a $G/M/1/G$ queue, the loss ratio under the \textit{EDF-EDT} scheduling policy is less than that of  \textit{EDF-EAC} scheduling policy, i.e., $\alpha_{EDF\mbox{-}EDT }^{G} \leq \alpha_{EDF\mbox{-}EAC }^{G}$.

\bigskip
\textit{Proposition} 7. 
	 In a $G/M/1/G$ queue, the loss ratio under the \textit{EDF-EDT} scheduling policy is less than that of  \textit{FCFS-EDT} scheduling policy, i.e., $\alpha_{EDF\mbox{-}EDT }^{G} \leq \alpha_{FCFS\mbox{-}EDT }^{G}$.

\bigskip	
\textit{Proposition} 8. 
	In a $G/G/1/G$ queue, the loss ratios under the \textit{FCFS-EDT} and \textit{FCFS-EAC} scheduling policies are identical, i.e., $\alpha_{FCFS\mbox{-}EAC }^{G} = \alpha_{FCFS\mbox{-}EDT }^{G}$.

\bigskip The following example shows that there is no dominance relation between the loss ratios under the \textit{EDF} and \textit{FCFS-EAC} (or \textit{FCFS-EDT}) scheduling policies, i.e., neither of the inequalities $\alpha_{FCFS\mbox{-}EAC}^{G} \le \alpha_{EDF}^{G}$ and $\alpha_{EDF}^{G} \le \alpha_{FCFS\mbox{-}EAC}^{G}$ hold in general.

\begin{figure}
	\centering
		\includegraphics[width=2.5in]{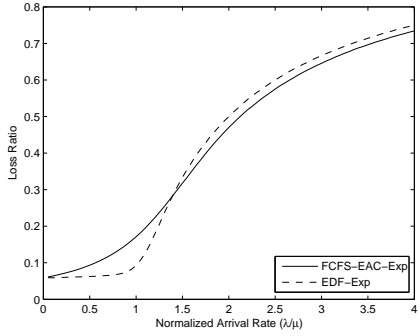}
		\caption{Loss ratios of the \textit{FCFS-EAC} and \textit{EDF} scheduling algorithms for exponential relative deadline with $1/\delta = 16/\mu$ and various normalized arrival rates ($\lambda/\mu$).}
		\label{fig:cteg_e}
\end{figure}

\bigskip 
	\textit{Counter-example 1.} Consider the $M/M/1$ queue with deadline till the end of the service, where the relative deadline has the exponential distribution with mean equal to 16 times the mean service time ($1/\delta=16/\mu$). The loss ratios, plotted in Figure~\ref{fig:cteg_e} as a function of the normalized arrival rate ($\lambda/\mu$), show that the inequality $\alpha_{EDF}^{Exp} \le \alpha_{FCFS\mbox{-}EAC}^{Exp}$ holds for small arrival rates, while the inequality $\alpha_{FCFS\mbox{-}EAC}^{Exp} \le \alpha_{EDF}^{Exp}$ holds for large arrival rates. The values of the loss ratios are computed on the basis of simulations of about one million arrivals. Thus, neither of $\alpha_{EDF}^{G}$ and $\alpha_{FCFS\mbox{-}EAC}^{G}$ uniformly dominates the other.$\hfill$

\begin{figure}
	\centering
		\includegraphics[width=2.5in]{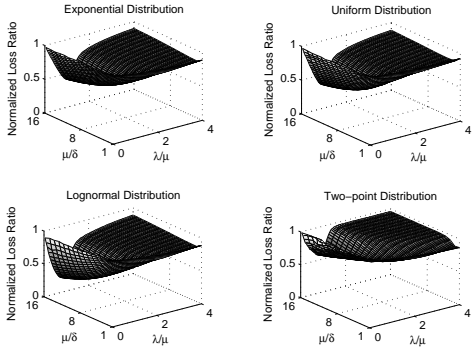}
		\caption{Loss ratios for various deadline distributions under the {\it EDF-EAC} scheduling policy normalized by loss ratio under the {\it FCFS-EAC} scheduling policy, for various values of normalized arrival rate ($\lambda/\mu$) and normalized mean relative deadline ($\mu/\delta$).}
		\label{fig:loss_ratio_Fs_Eac}
\end{figure}

The results of propositions~1 and~7 give rise to the question of possible superiority of EDF over FCFS in terms of loss ratio even under exact admission control. We could not prove this dominance relation. However, simulation results summarized in Figure~\ref{fig:loss_ratio_Fs_Eac} appear to support this. We ran the simulations for the Poisson arrival process with a wide range of normalized arrival rates (with $\lambda/\mu$ varying from 0 to 4) and four types of relative deadline distributions, namely exponential, uniform, log-normal and two-point. The mean ($1/\delta$) of the deadline distribution was varied from $1/\mu$ to $16/\mu$. In the case of the log-normal distribution, the coefficient of variation was chosen as 1 for all values of $\delta$, while in the case of the two-point distribution, the probabilities 0.9 and 0.1 were assigned to the points $5/(9\delta)$ and $5/\delta$, respectively, for all values of $\delta$. The values of the loss ratios were computed on the basis of simulations of about one million arrivals. On the basis of these findings, we make the following conjecture.

\bigskip	
\textit{Conjecture} 1. 
	 In an $G/M/1/G$ queue, the loss ratio under the \textit{EDF-EAC} scheduling policy is less than that of  \textit{FCFS-EAC} scheduling policy, i.e., $\alpha_{EDF\mbox{-}EAC}^{G} \leq \alpha_{FCFS\mbox{-}EAC}^{G}$.
	 
\begin{figure}
	\centering
		\includegraphics[width=2.5in]{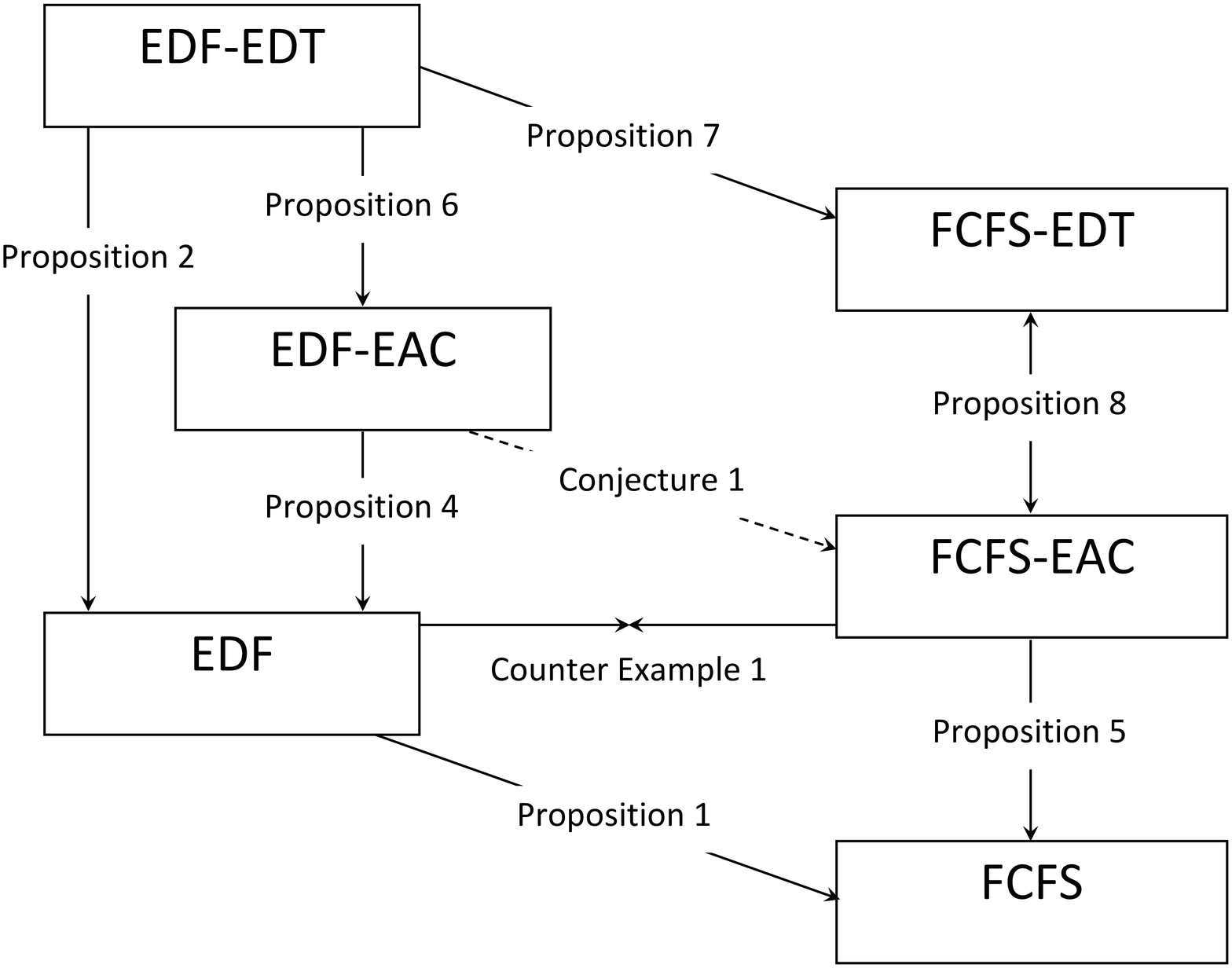}
		\caption{Relationship between various scheduling algorithms in terms of order of loss ratios, for stochastic relative deadlines.}
		\label{fig:relationship_diagram1}
\end{figure}

\bigskip	
The findings of this section are summarized in Figure~\ref{fig:relationship_diagram1}. In this figure, an arrow extending from the scheduling policy $sp_1$ to the policy $sp_2$ indicates that $\alpha_{sp_1 }^{G}\le\alpha_{sp_2 }^{G}$, a double headed arrow indicates equality of the loss ratios, while a pair of arrows facing each other indicates that there is no dominance relation. The dashed arrow represents a conjectured relation.

\bigskip	
A key finding that emerges from Figure~\ref{fig:relationship_diagram1} is the superiority of {\it EDF-EDT} over the other scheduling policies, in terms of loss ratio.

\section{Deterministic deadline} \label{degenerate}
\begin{figure}
	\centering
		\includegraphics[width=2.5in]{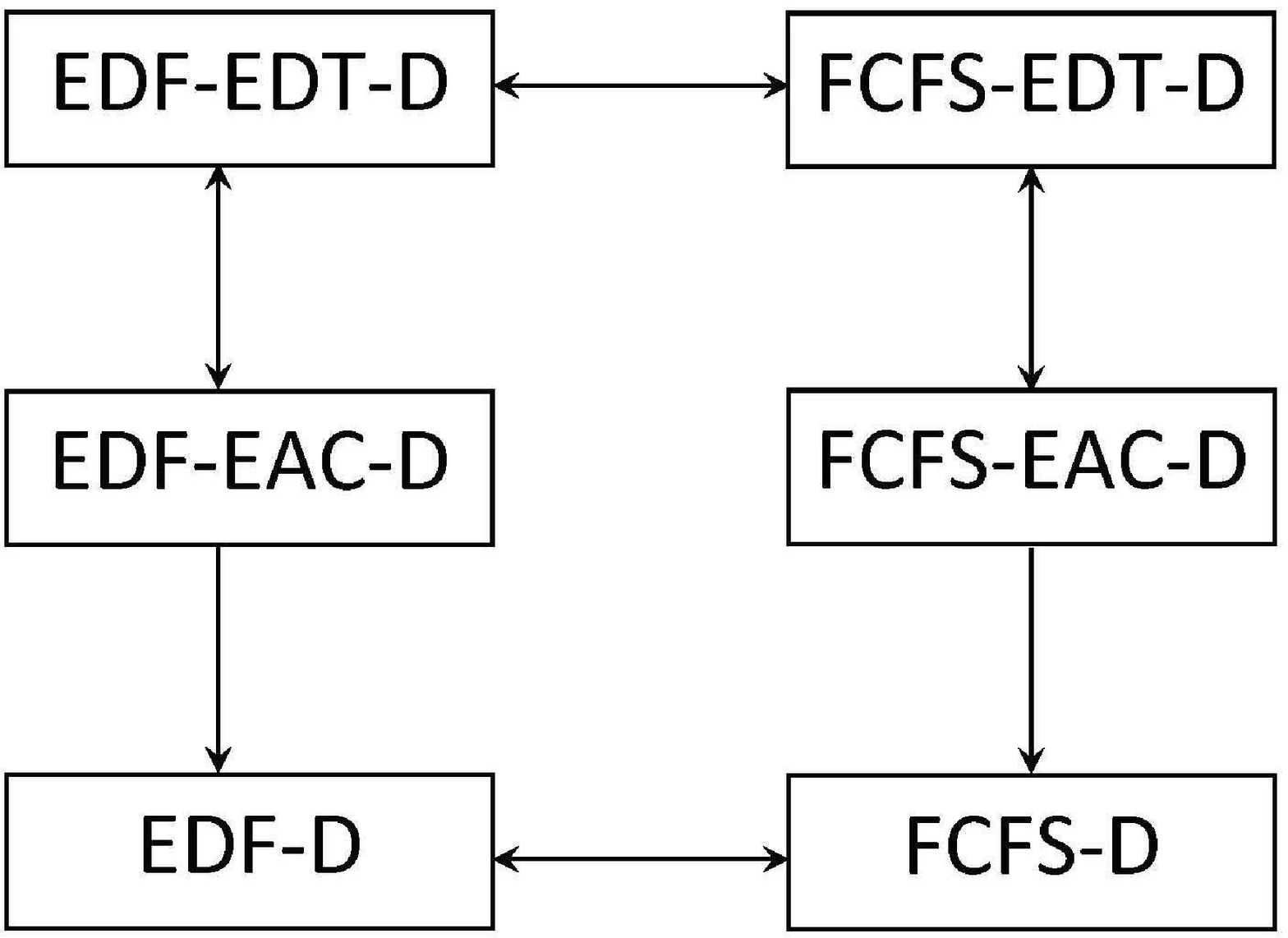}
		\caption{Relationship between various scheduling algorithms in terms of order of loss ratios, for deterministic relative deadlines.}
		\label{fig:relationship_diagram2}
\end{figure}

We now assume that the deadline distribution is degenerate, i.e., the deadline is completely deterministic. It is easy to see the following facts. 
\begin{enumerate}
\item The \textit{FCFS} and \textit{EDF} scheduling policies are equivalent.
\item The \textit{FCFS-EDT} and \textit{EDF-EDT} scheduling policies are equivalent.
\item The \textit{FCFS-EDT} and \textit{FCFS-EAC} scheduling policies are equivalent.
\item The \textit{EDF-EDT} and \textit{EDF-EAC} scheduling policies are equivalent.
\end{enumerate}
In view of the above facts, the relations depicted in Figure~\ref{fig:relationship_diagram1} simplify to those given in Figure~\ref{fig:relationship_diagram2}.

\bigskip
Let $\alpha_{sp }^{Det}$	denote the loss ratio of the system under scheduling policy \textit{sp} and deterministic relative deadline. Movaghar \cite{mova_des} showed that the loss ratio for the FCFS scheduling policy is bounded from below by the corresponding ratio for the case where the deadline is deterministic. In particular, the following proposition follows from Lemma~5.1.3 of Movaghar \cite{mova_des}.

\bigskip
\textit{Proposition} 9. 
	In an $M/M/1/G$ queue with a specified mean deadline till the end of service, the loss ratio under the \textit{FCFS} scheduling policy happens to be the minimum when the deadline distribution is degenerate, i.e., $\alpha_{FCFS}^{Det}\le\alpha_{FCFS}^{G}$.

\bigskip
The above result provides a connection between two loss ratios shown in Figures~\ref{fig:relationship_diagram1} and~\ref{fig:relationship_diagram2}. Questions about similar other connections arise naturally. An optimality result of the type of Proposition~9 for the EDF scheduling policy was conjectured in \cite{mom}, and we state it below.

\bigskip
\textit{Conjecture} 2. 
In an $M/M/1/G$ queue with a specified mean deadline till the end of service, the loss ratio under the {\it EDF} scheduling policy happens to be the minimum when the deadline distribution is degenerate, i.e., $\alpha_{EDF}^{Det}\le\alpha_{EDF}^{G}$.

\begin{figure}
	\centering
		\includegraphics[width=2.5in]{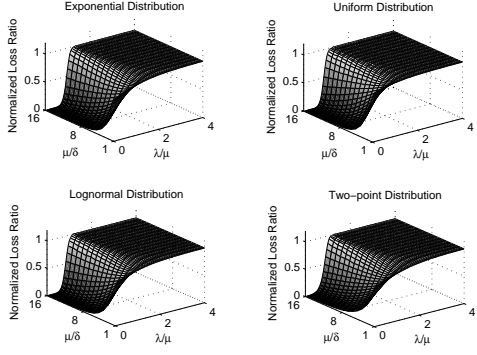}
		\caption{Loss ratio for deterministic deadline normalized by loss ratios for various deadline distributions under the {\it EDF} scheduling policy, for various values of normalized arrival rate ($\lambda/\mu$) and normalized mean relative deadline ($\mu/\delta$).}
		\label{fig:loss_ratio_GD1}
\end{figure}

\bigskip
We were unable to find either a proof of the above conjecture or a counter-example to disprove it. However, we conducted extensive simulations for a number of relative deadline distributions. We considered Poisson arrival process with a wide range of normalized arrival rates (with $\lambda/\mu$ varying from 0 to 4), and four types of relative deadline distributions, namely exponential, uniform, log-normal and two-point. The mean ($1/\delta$) of the deadline distribution was varied from $1/\mu$ to $16/\mu$. As in the case of simulations run for {\it Conjecture 1}, the coefficient of variation of the log-normal distribution was fixed as 1 for all values of $\delta$, and the points $5/(9\delta)$ and $5/\delta$ of the two point distribution were assigned probabilities 0.9 and 0.1, respectively, for all values of $\delta$. The values of the loss ratio were computed on the basis of simulations of about one million arrivals. The results, summarized in Figure~\ref{fig:loss_ratio_GD1}, support the above conjecture.

\begin{figure}
	\centering
		\includegraphics[width=2.5in]{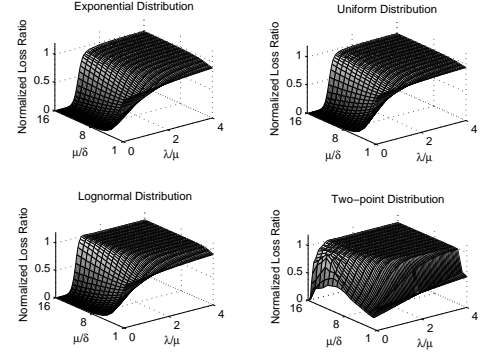}
		\caption{Loss ratio for deterministic deadline normalized by loss ratios for various deadline distributions under the {\it EDF-EDT} scheduling policy, for various values of normalized arrival rate ($\lambda/\mu$) and normalized mean relative deadline ($\mu/\delta$).}
		\label{fig:loss_ratio_GD2}
\end{figure}

We looked for a similar result for the EDF-EDT scheduling policy, but were unable to find either a proof or a counter-example. We state it in the form of a conjecture, which is supported by the simulation results summarized in Figure~\ref{fig:loss_ratio_GD2}. The model of this experiment is the same as before.

\bigskip
\textit{Conjecture} 3. 
In an $M/M/1/G$ queue with a specified mean deadline till the end of service, the loss ratio under the {\it EDF-EDT} scheduling policy happens to be the minimum when the deadline distribution is degenerate, i.e., $\alpha_{EDF-EDT}^{Det}\le\alpha_{EDF-EDT}^{G}$.

\bigskip
The following three counter-examples complete the set of connections between the loss ratios depicted in Figures~\ref{fig:relationship_diagram1} and~\ref{fig:relationship_diagram2}.
	
\begin{figure}
	\centering
		\includegraphics[width=2.5in]{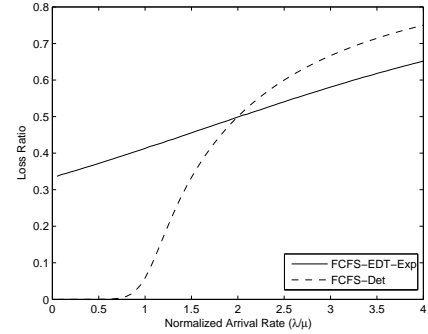}
		\caption{Loss ratios of the \textit{FCFS-EDT,Exp} and \textit{FCFS,Det} scheduling algorithms for mean relative deadline $1/\delta = 2/\mu$ and various normalized arrival rates ($\lambda/\mu$).}		
		\label{fig:cteg_ed}
\end{figure}

\bigskip
	\textit{Counter-example 2.} Consider the $M/M/1$ queue with deadline till the end of the service, where the relative deadline is either deterministic with value $2/\mu$ or exponentially distributed with mean $2/\mu$. Loss ratios plotted in Figure~\ref{fig:cteg_ed} as a function of the normalized arrival rate ($\lambda/\mu$), show that the inequality $\alpha_{FCFS}^{Det} \le \alpha_{FCFS\mbox{-}EDT}^{Exp}$ holds for small arrival rates, while the inequality $\alpha_{FCFS\mbox{-}EDT}^{Exp} \le \alpha_{FCFS}^{Det}$ holds for large arrival rates. The values of the loss ratio were computed on the basis of simulations of about one million arrivals. Thus, neither of $\alpha_{FCFS}^{Det}$ and $\alpha_{FCFS\mbox{-}EDT}^{G}$ uniformly dominates the other.$\hfill$
\bigskip

\begin{figure}
	\centering
		\includegraphics[width=2.5in]{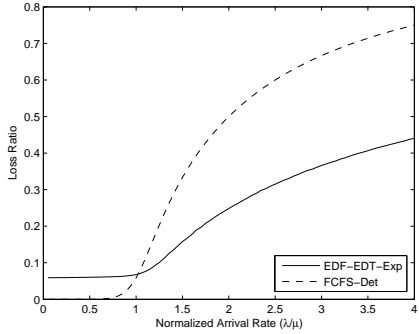}
		\caption{Loss ratios of the \textit{EDF-EDT,Exp} and \textit{FCFS,Det} scheduling algorithms for mean relative deadline $1/\delta=16/\mu$ and various normalized arrival rates ($\lambda/\mu$).}
		
		\label{fig:cteg_efd}
\end{figure}

	\textit{Counter-example 3.} Consider the $M/M/1$ queue with deadline till the end of the service, where the relative deadline is either deterministic with value $16/\mu$ or exponentially distributed with mean $16/\mu$. The loss ratios, plotted in Figure~\ref{fig:cteg_efd} as a function of the normalized arrival rate ($\lambda/\mu$), show that the inequality $\alpha_{FCFS }^{Det} \le \alpha_{EDF\mbox{-}EDT }^{Exp}$ holds for small arrival rates, while the inequality $\alpha_{EDF\mbox{-}EDT }^{Exp} \le \alpha_{FCFS }^{Det}$ holds for large arrival rates. The values of the loss ratio were computed on the basis of simulations of about one million arrivals. Thus, neither of $\alpha_{FCFS }^{Det}$ and $\alpha_{EDF\mbox{-}EDT }^{G}$ uniformly dominates the other.$\hfill$
\bigskip

\begin{figure}
	\centering
		\includegraphics[width=2.5in]{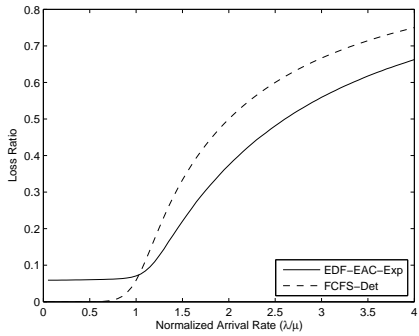}
		\caption{Loss ratios of the \textit{EDF-EAC,Exp} and \textit{FCFS,Det} scheduling algorithms for mean relative deadline $1/\delta=16/\mu$ and various normalized arrival rates ($\lambda/\mu$).}
		\label{fig:cteg_efa}
\end{figure}

	\textit{Counter-example 4.} Consider the $M/M/1$ queue with deadline till the end of the service, where the relative deadline is either deterministic with value $16/\mu$ or exponentially distributed with mean $16/\mu$. The loss ratios, plotted in Figure~\ref{fig:cteg_efa} as a function of the normalized arrival rate ($\lambda/\mu$), show that the inequality $\alpha_{FCFS }^{Det} \le \alpha_{EDF\mbox{-}EAC }^{Exp}$ holds for small arrival rates, while the inequality $\alpha_{EDF\mbox{-}EAC }^{Exp} \le \alpha_{FCFS }^{Det}$ holds for large arrival rates. The values of the loss ratio were computed on the basis of simulations of about one million arrivals. Thus, neither of $\alpha_{FCFS }^{Det}$ and $\alpha_{EDF\mbox{-}EAC }^{G}$ uniformly dominates the other.$\hfill$

\section{Concluding Remarks} \label{conclusion}
\begin{figure}
	\centering
		\includegraphics[width=2.5in]{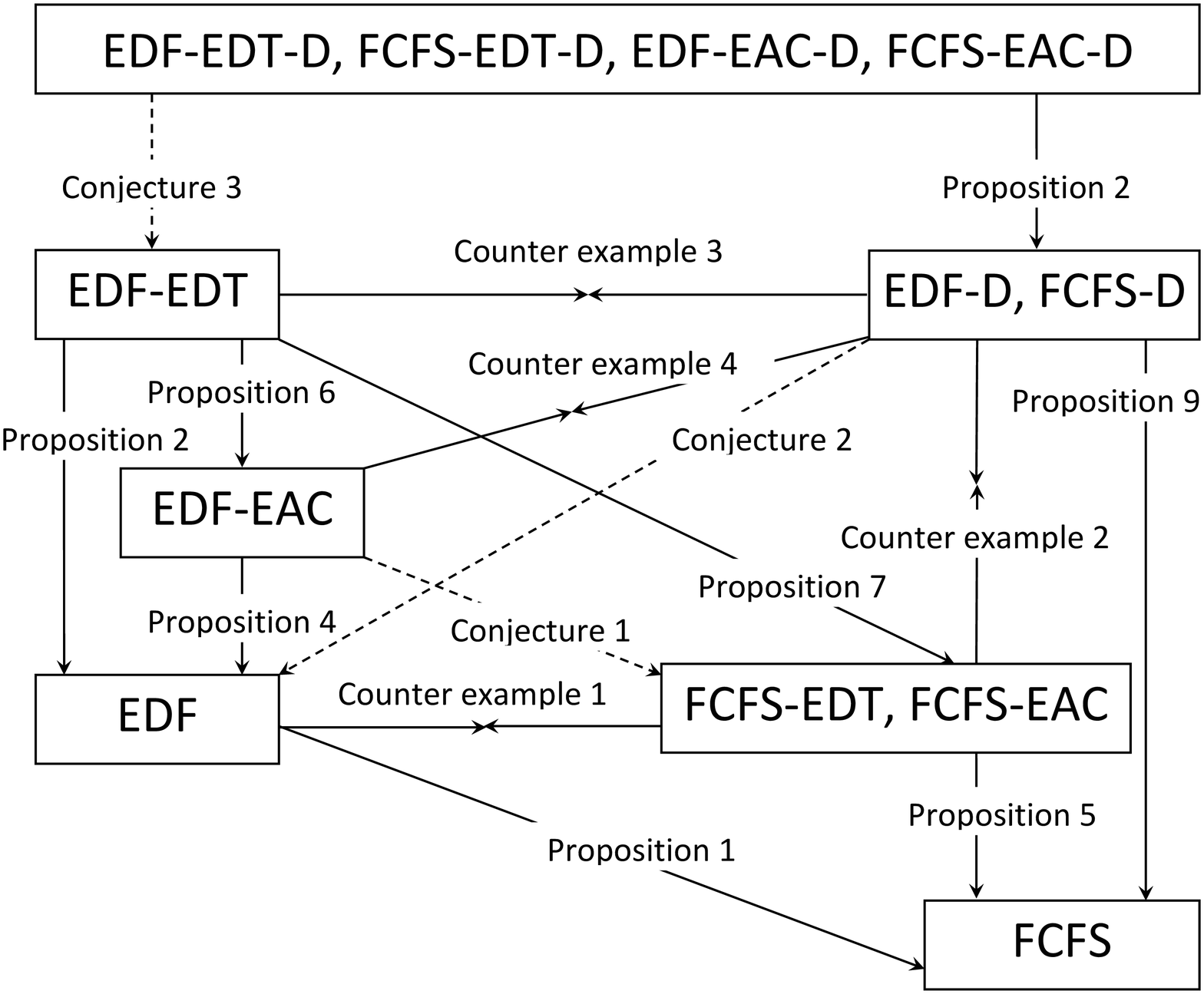}
		\caption{Relationship between various scheduling algorithms in terms of order of loss ratios, for stochastic and deterministic relative deadlines.}
		\label{fig:relationship_diagram3}
\end{figure}

In this paper, we have proved some dominance relations between various scheduling algorithms in terms of their respective loss ratios. We have also proved, through counter-examples, the non-existence of a dominance relation between some pairs of scheduling algorithms. A few possible dominance relations are left as conjectures, supported by extensive simulations. These relations help one construct a comprehensive dominance structure of scheduling algorithms in terms of loss ratios, parts of which were given in Figures~\ref{fig:relationship_diagram1} and~\ref{fig:relationship_diagram2}. This combined structure is shown in Figure~\ref{fig:relationship_diagram3}.

An intuitive explanation of the smaller loss ratio of EDF-EDT in comparison to EDF-EAC is the fact that, while taking the decision regarding admission of a job, an admission controller takes into account the history of jobs {\it already in the queue}. Some of these decisions may appear to be unduly conservative in the light of events that follow. By deferring the decision of discarding a job till the epoch of its being served, EDT is able to take into account additional information.

The result of Proposition~6 indicates that EDF-EDT may be preferred where an early guarantee of completion of jobs is not essential. On the other hand, if an admission controller must be used, then Propositions~4 and~6 specify limits to its performance from both sides. 

The result of Towsley and Panwar \cite{opti_des} on the optimality of the EDF scheduling policy among the class of all service-time independent policies does not hold in the presence of EDT, which makes the scheduling policy dependent on service time. This fact gives rise to the question of possible optimality of EDF among the modified class of scheduling policies that accommodate EDT. The result stated in Proposition~7 partially answers that question, and keeps open the possibility of overall optimality. This issue may be taken up for research in future.

\appendices
\section{Proofs of propositions}
\textit{Proof of Proposition 2.} Consider a finite number of job arrivals, and arrange all the jobs in order of their departure epochs under \textit{EDF}. Observe that the departure order of \textit{EDF-EDT} is the same as that under \textit{EDF}. A job that is discarded under EDT would have missed the deadline in any case. On the other hand, the act of discarding a particular job can only reduce the waiting times of the subsequent jobs (arranged as above). Consequently, the act of discarding that job can only reduce the number of subsequent jobs missing their respective deadlines. This argument holds for every single event of discarding of jobs under EDT. Thus, for any given configuration of a finite number of jobs, the proportion of jobs missing deadline under \textit{EDF-EDT} is less than or equal to that under \textit{EDF}. It follows that the expected proportion of jobs (out of the first $n$ arrivals for any fixed $n$) is less for \textit{EDF-EDT} than for \textit{EDF}. The stated result is obtained by taking the limit of the expected proportions as $n$ goes to infinity.$\hfill$ 
	 
	\bigskip\textit{Proof of Proposition 3.} The result can be proved along the lines of the proof of Proposition 2.$\hfill$
	 
	\bigskip\textit{Proof of Proposition 4.} \cite{sup_edt_over_eac} Consider a variation of the \textit{EDF-EAC} scheduling policy, where a job that does not satisfy the admission criterion of \textit{EAC} is not discarded at the time of admission, rather it is merely tagged for eventual rejection at the epoch of its getting the server. Note that this modification does not change the completion status of any job, but the order of the jobs (tagged or untagged) getting the server becomes the same as that under \textit{EDF}. We shall show that the modified \textit{EDF-EAC} procedure produces a smaller loss ratio than \textit{EDF}. 
	
	 Now, consider the first $n$ arrivals. Permit the first tagged job of the list to be served, allowing for possible preemption under \textit{EDF}. The fact that the tagged job would have been denied admission under \textit{EAC} indicates that either this job or at least one job that arrived earlier but is located down the queue would miss deadline. On the other hand, providing service to the tagged job can only increase the waiting times of \textit{all} the subsequent jobs in the ordered list (including those which arrived after the first tagged job), and this increase may trigger further cases of missed deadline. Thus, the act of providing service to the tagged job can only increase the number of jobs missing deadline. This argument holds for every single event of providing service to the successively tagged jobs. Thus, for any given configuration of a finite number of jobs, the proportion of jobs missing deadline under \textit{EDF-EAC} is less than or equal to that under \textit{EDF}. Hence, the expected proportion of jobs (out of the first $n$ arrivals) is less for \textit{EDF-EDT} than for \textit{EDF}. The stated result is obtained by taking the limit of the expected proportion as $n$ goes to infinity.$\hfill$ 
	 
	\bigskip\textit{Proof of Proposition 5.} The result can be proved along the lines of the proof of Proposition 4.$\hfill$

	\bigskip\textit{Proof of Proposition 6.} \cite{sup_edt_over_eac} Consider the task of scheduling the first $N$ jobs in a $G/M/1/G$ queue. For $n=0,1,2,\ldots,N$, let $P_n$ denote the scheduling policy, where the jobs are scheduled according to the \textit{EDF-EAC} policy for the first $n$ arrivals, and there is a switch to the \textit{EDF-EDT} policy before the arrival of the next job. Note that $P_0$ corresponds to the \textit{EDF-EDT} policy, while $P_N$ corresponds to the \textit{EDF-EAC} policy. If we can show that the expected count of completed jobs is a decreasing function of $n$, then the stated result will follow by allowing $N$ to go to infinity.
	
	In order to compare the policies $P_{n-1}$ and $P_n$, consider three cases depending on the admission status of the job $J$ that corresponds to the $n$th arrival.
	
	{\sc Case 1.} Let $J$ be admissible according to \textit{EAC}. In this case, the pattern of service provided under the policies $P_{n-1}$ and $P_n$ would be identical.

	{\sc Case 2.} Let $J$ be inadmissible according to \textit{EAC} owing to the fact that its own deadline is too short for its completion. In this case also, the pattern of service provided under the policies $P_{n-1}$ and $P_n$ would be identical. The only difference is that $P_n$ would not admit $J$, while $P_{n-1}$ would admit it but discard it at the epoch of its getting server.

	{\sc Case 3.} Let $J$ be inadmissible according to \textit{EAC} because of the fact that its admission would result in non-completion of service to another job. Consider the entire history of arrivals and server utilization subsequent to the arrival of $J$. Let $J'$ be the label of the first job (arrived before or after $J$) that fails to complete because of the admission of $J$ under $P_{n-1}$. Let us denote by $J^*$ the job immediately preceding $J'$ in the queue according to $P_n$. Regarding the arrival epoch of $J$ as time 0, let $\tau$ and $\tau+\tau^*$ be the remaining aggregated service times of the jobs having absolute deadlines earlier than those of $J$ and $J'$, respectively. Let $X$ and $X'$ be the remaining service times of $J$ and $J'$, respectively, at time 0. Let $d$, $d+d^*$ and $d+d^*+d'$ be the absolute deadlines of $J$, $J^*$ and $J'$, respectively. In case $J'$ immediately follows $J$ in the queue of $P_{n-1}$ (i.e., there is no job with the label $J^*$), we set $\tau^*$ and $d^*$ as zero.
	
We shall show that, given these circumstances, $X$ is stochastically smaller than $X'$. 

For any set of fixed and positive values of $\tau$, $\tau^*$, $d$, $d^*$ and $d'$ satisfying $\tau<d$ and $\tau+\tau^*<d+d^*$, consider the event
\begin{eqnarray*}
E&=&\{\tau+X\le d;\ \tau+X+\tau^*\le d+d^*;\\\ &&\tau+\tau^*+X'\le d+d^*+d'<\tau+X+\tau^*+X'\}.
\end{eqnarray*}

Note that this event represents the conditions that $J$ and $J^*$ can be completed before their respective deadlines according to $P_{n-1}$, while $J'$ can be completed under $P_n$ but not under $P_{n-1}$.	If the service rate is $\mu$, it is easy to see that the joint density of $X$ and $X'$ given $E$ is 
$$f_{X,X'|E}(x,x')=\frac{\mu^2e^{-\mu(x+x')}}{P(E)},\quad x\le a,\ x'\le b<x+x',$$
where $a=\min\{d-\tau,d+d^*-\tau-\tau^*\}$, $b=d+d^*+d'-\tau-\tau^*$ (with $a\le b$)
and $P(E)$ is the unconditional probability
$$\mathop{\int\int}_{x\le a,\ x'\le b<x+x'}\mu^2e^{-\mu(x+x')}\,dx\,dx'.$$
Now, it follows that the conditional density of $X$ given $E$ is
\begin{eqnarray*}
f_{X|E}(x)&=&\int_0^\infty f_{X,X'|E}(x,x')\,dx'\\
&=&\int_{b-x}^{b}\frac{\mu^2e^{-\mu(x+x')}}{P(E)}\,dx'\\
&=&\frac{\mu}{P(E)}e^{-\mu b}\left(1-e^{-\mu x}\right),\quad 0\le x\le a,
\end{eqnarray*}
and the corresponding distribution function is
$$F_{X|E}(x)=
\left\{\begin{array}{ll}0&\mbox{if }x\le0,\\
\displaystyle{\frac{e^{-\mu b}}{P(E)}\left[\mu x - \left(1-e^{-\mu x}\right)\right]}
&\mbox{if }0<x\le a,\\
1&\mbox{if }x>a.\\ \end{array}\right.$$
On the other hand, the conditional density of $X'$ given $E$ is
\begin{eqnarray*}
f_{X'|E}(x')\!\!
&=&\!\!\int_0^\infty f_{X,X'|E}(x,x')\,dx\\
&=&\!\!\int_{b-x'}^{a}\frac{\mu^2e^{-\mu(x+x')}}{P(E)}\,dx\\
&=&\!\!\frac{\mu}{P(E)}\left[e^{-\mu b}-e^{-\mu(x'+a)}\right],\quad b-a\le x'\le b,
\end{eqnarray*}
and the corresponding distribution function is
\begin{eqnarray*}
&&\hskip-25pt F_{X'|E}(x')\\
&=&\!\!
\left\{\begin{array}{l@{\hskip-70pt}l}0&\mbox{if }x'\le b-a,\\[2ex]
\displaystyle{\frac{e^{-\mu b}}{P(E)}\left[\mu\left\{ x' - (b-a)\right\} -\left\{1-e^{-\mu\left\{ x' - (b-a)\right\}}\right\}\right]}\\
&\mbox{if }b-a< x'\le b,\\[2ex]
1\\&\mbox{if }x'>b. \end{array}\right.
\end{eqnarray*}
By comparing the conditional distribution functions of $X$ and $X'$, we observe that the inequality $F_{X|E}(t)\ge F_{X'|E}(t)$ holds trivially in the ranges $t\le b - a$ and $t>a$. For the remaining range of values of $t$ (if such values are feasible, i.e., if $b-a<a$), the inequality holds if and only if
$$\mu (b-a)\ge e^{-\mu t}\left(e^{\mu (b-a)}-1\right)$$
for $b-a<t\le a$. The right hand side assumes its largest value for $t=b-a$, and hence a sufficient condition for the above inequality to hold is $\mu (b-a)\ge e^{-\mu (b-a)}\left(e^{\mu (b-a)}-1\right)$, i.e., $e^{-\mu (b-a)}\ge 1-\mu (b-a)$. The last condition is always satisfied. This proves that $X'$ is stochastically larger than $X$.

The total count of completed jobs up to the disposing of $J'$ is the same under $P_n$ and $P_{n-1}$. The expected number of jobs completed after the disposing of $J'$ is a decreasing function of the time of disposal of $J'$. Note that, in a time scale that starts from the arrival epoch of $J$, this disposal time is either $\tau+X+\tau^*$ or $\tau+\tau^*+X'$, depending on whether $P_{n-1}$ or $P_n$ is used. Since $X$ is stochastically smaller than $X'$, the expected total number of completed jobs is larger under $P_{n-1}$ than under $P_n$. 

After combining the findings of the three cases, we observe that the expected proportion of completed jobs (out of $N$ arrived jobs) under the scheduling policy $P_n$ is a decreasing function of $n$. This completes the proof.$\hfill$

	\bigskip\textit{Proof of Proposition 7.} \cite{sup_edfedt_over_fcfsedt} As in the proof of Proposition~2.6, consider the task of scheduling the first $N$ jobs. For $n=0,1,2,\ldots,N$, let $P_n$ denote the scheduling policy, where the jobs are scheduled according to the \textit{EDF-EDT} policy for the first $n$ arrivals, and there is a switch to the \textit{FCFS-EDT} policy before the arrival of the next job. Note that $P_0$ corresponds to the \textit{FCFS-EDT} policy, while $P_N$ corresponds to the \textit{EDF-EDT} policy. We aim at showing that the expected count of completed jobs is an increasing function of $n$.
	
	Once again, we consider three cases depending on the status of the job $J$ that corresponds to the $n$th arrival.	
	
	{\sc Case 1.} Let $J$ be successfully serviceable as of the time of its arrival, if it is scheduled according to \textit{FCFS-EDT}. It follows that it is successfully serviceable as of the time of its arrival, if it is scheduled according to \textit{EDF-EDT} also. In this case, the completion status of all the jobs under the policies $P_{n-1}$ and $P_n$ would be identical, even though the two policies may place $J$ in different positions of the queue.
	
	{\sc Case 2.} Let $J$ not be successfully serviceable as of the time of its arrival, if it is queued according to \textit{EDF-EDT}. It follows that it is not successfully serviceable as of the time of its arrival, if it is scheduled according to \textit{FCFS-EDT} also. In this case also, the completion status of all the jobs under the policies $P_{n-1}$ and $P_n$ would be identical.
	
	{\sc Case 3.} Let $J$ be successfully serviceable as of the time of its arrival, if it is queued according to \textit{EDF-EDT} but not so under \textit{FCFS-EDT}. Let $J'$ be the label of the first job (arrived before or after $J$) that fails to complete after being superseded by $J$ under $P_n$. Let $X$ and $X'$ be the remaining service times of $J$ and $J'$, respectively, at time 0. It can be proved along the lines of the proof of Proposition 2.6 that $X$ is stochastically smaller than $X'$, and that the expected total number of completed jobs is smaller under $P_{n-1}$ than under $P_n$. 

	The proof is completed by combining the findings of the three cases for fixed $N$ (which establishes that $P_0$ has  smaller loss ratio than $P_N$), and then allowing $N$ to go to infinity.$\hfill$

	\bigskip\textit{Proof of Proposition 8.} Consider a variation of the \textit{FCFS-EAC} scheduling policy, where a job that does not satisfy the admission criterion of \textit{EAC} is not discarded at the time of admission, rather it is merely tagged for eventual rejection at the epoch of its getting the server. Note that this modification does not change the completion status of any job, but the order of the jobs (tagged or untagged) getting the server becomes the same as that under \textit{FCFS}.	The fact that the tagged job would have been denied admission under the \textit{FCFS-EAC} procedure indicates that this job, if served, would have missed its own deadline. Thus, this job would also be discarded under \textit{FCFS-EDT}. It can be seen that, out of the first $n$ arrivals, the set of jobs that would be discarded under \textit{FCFS-EDT} is precisely the set of jobs tagged as above. It follows that, for any given configuration of $n$ job arrivals, the proportion of jobs missing deadline is the same under \textit{FCFS-EAC} and \textit{FCFS-EDT}. The result follows by taking expectation of this proportion and allowing $n$ to go to infinity.$\hfill$
	

%
%

%

%
%
%
%
%
\end{document}